\documentclass[reprint,superscriptaddress,longbibliography, prd]{revtex4-1}
\usepackage{color}
\usepackage[normalem]{ulem}
\usepackage{graphicx}
\usepackage{hyperref}
\usepackage{amsmath}
\usepackage{amssymb}
\usepackage{array}
\usepackage{dcolumn}
\usepackage{color}
\usepackage{bm}
\usepackage{multirow}
\usepackage{makecell}
\usepackage{subfigure}
\usepackage{soul}
\usepackage{tabularx}
\newcolumntype{C}{>{\centering\arraybackslash}p{3.8cm}}

\hypersetup{
    pdfnewwindow=true,    
    colorlinks=true,      
    linkcolor=blue,       
    citecolor=blue,       
    filecolor=blue,       
    urlcolor=blue         
}

\definecolor{forestgreen}{rgb}{0.13, 0.55, 0.13}

\begin{document}

\title{Betelgeuse Constraints on Coupling between Axion-like Particles and Electrons}

\author{Mengjiao Xiao}
    \email{mjxiao@mit.edu}
    \affiliation{Department of Physics, Massachusetts Institute of Technology, Cambridge, Massachusetts 02139, USA}
\author{Pierluca Carenza}
    \email{pierluca.carenza@fysik.su.se}
    \affiliation{The Oskar Klein Centre, Department of Physics, Stockholm University, Stockholm 106 91, Sweden}
\author{Maurizio Giannotti}
    \email{mgiannotti@barry.edu}
    \affiliation{Physical Sciences, Barry University, 11300 NE 2nd Ave., Miami Shores, Florida 33161, USA}
\author{Alessandro Mirizzi}
    \email{alessandro.mirizzi@ba.infn.it}
    \affiliation{Dipartimento Interateneo di Fisica “Michelangelo Merlin”, Via Amendola 173, 70126 Bari, Italy}
    \affiliation{Istituto Nazionale di Fisica Nucleare - Sezione di Bari, Via Orabona 4, 70126 Bari, Italy}
\author{Kerstin M. Perez}
    \email{kmperez@mit.edu}
    \affiliation{Department of Physics, Massachusetts Institute of Technology, Cambridge, Massachusetts 02139, USA}
\author{Oscar Straniero}
    \email{oscar.straniero@inaf.it}
    \affiliation{INAF, Osservatorio Astronomico d’Abruzzo, 64100 Teramo, Italy}
\author{Brian W. Grefenstette}
    \email{bwgref@srl.caltech.edu}
    \affiliation{Cahill Center for Astrophysics, 1216 E. California Blvd, California Institute of Technology, Pasadena, California 91125, USA}
\date{\today}

\begin{abstract}
Axion-like particles (ALPs) can be produced by thermal processes in a stellar interior, escape from the star and, if sufficiently light, be converted into photons in the external Galactic magnetic field. Such a process could produce a detectable hard X-ray excess in the direction of the star. In this scenario, a promising class of targets is the red supergiants, massive stars which are experiencing the late part of their evolution. We report on a search for ALP-induced X-ray emission from Betelgeuse, produced via the combined processes of Bremsstrahlung, Compton and Primakoff. Using a 50\,ks observation of Betelgeuse by the \emph{NuSTAR} satellite telescope, we set 95\% C.L. upper limits on the ALP-electron ($g_{ae}$) and ALP-photon ($g_{a\gamma}$) couplings. For masses ${m_{a}\leq(3.5-5.5)\times10^{-11}}$ eV, we find
$g_{a\gamma} \times g_{ae}< (0.4-2.8)\times10^{-24}$ \textrm{GeV}$^{-1}$ (depending on the stellar model and assuming a value of the regular Galactic magnetic field in the direction transverse to Betelgeuse of $B_T$=1.4~\,$\mu$G). This corresponds to ${g_{ae}<(0.4-2.8) \times10^{-12}}$ for
${g_{a\gamma}>1.0\times10^{-12}}$ \textrm{GeV}$^{-1}$. This analysis supercedes by over an order of magnitude the limit on $g_{ae} \times g_{a\gamma}$ placed by the CAST solar axion experiment and is among the strongest constraints on these couplings.
\end{abstract}

\maketitle

\nocite{*}

\begin{table*}[ht!]
\centering
\renewcommand{\arraystretch}{1.35}
\begin{tabular*}{2.0\columnwidth}{@{\extracolsep{\fill}}ccccc|ccc|ccc|ccc}
    \hline\hline
    \multirow{2}*{Model} & \multirow{2}*{Phase} & \multirow{2}*{$t_\mathrm{cc}$ [yr]} & \multirow{2}*{$\log_{10}\frac{L_\mathrm{eff}}{L_\odot}$} & 
    \multirow{2}*{$\log_{10}\frac{T_\mathrm{eff}}{ \mathrm{K}}$} & \multicolumn{3}{c|}{Primakoff} & \multicolumn{3}{c|}{Bremsstrahlung} & \multicolumn{3}{c}{Compton} \\
    \cline{6-14}
    & & & & & $C^{P}$ & $E_{0}^{P}$ [keV] &$\beta^{P}$ & $C^{B}$ & $E_{0}^{B}$ [keV] &$\beta^{B}$ &
    $C^{C}$ & $E_{0}^{C}$ [keV] &$\beta^{C}$ \\
    \hline
    0  & He burning & 155000 & 4.90 & 3.572 & 1.36 & 50 & 1.95 &1.3E-3 &35.26 &1.16 &1.39 &77.86 &3.15 \\
    1 & before C burning & 23000 & 5.06 & 3.552 & 4.0 & 80 & 2.0 &2.3E-2 &56.57 &1.16 &8.55 &125.8 &3.12 \\
    2 & before C burning & 13000 & 5.06 & 3.552 & 5.2 & 99 & 2.0 &6.4E-2 &70.77 &1.09 & 17.39 &156.9 &3.09 \\
    3 & before C burning & 10000 & 5.09 & 3.549 & 5.7 & 110 & 2.0 &8.9E-2 &76.65 &1.08 &22.49 &169.2 &3.09 \\
    4 & before C burning & 6900  & 5.12 & 3.546 & 6.5 & 120 & 2.0 &0.136 &85.15 &1.06 &31.81 &186.4 &3.09 \\
    5 & in C burning & 3700  & 5.14 & 3.544 & 7.9 & 130 & 2.0 &0.249 &97.44 &1.04 &50.62 &210.4 &3.11\\
    6 & in C burning & 730   & 5.16 & 3.542 & 12 & 170 & 2.0 &0.827 &129.17 &1.02 &138.6 &269.1 &3.17 \\
    7 & in C burning & 480   & 5.16 & 3.542 & 13 & 180 & 2.0 &0.789 &134.54 &1.02 &153.2 &279.9 &3.15\\
    8 & in C burning & 110   & 5.16 & 3.542 & 16 & 210 & 2.0 &1.79 &151.46 &1.02 &252.7 &316.8 &3.17 \\
    9 & in C burning & 34    & 5.16 & 3.542 & 21 & 240 & 2.0 &2.82 &181.74 &1.00 &447.5 &363.3 &3.22 \\
    10 & between C/Ne burning & 7.2   & 5.16 & 3.542 & 28 & 280 & 2.0 &3.77 &207.84 &0.99 &729.2 &415.7 &3.23 \\
    11 & in Ne burning & 3.6 & 5.16 & 3.542 & 26 & 320 & 1.8 &3.86 &224.45 &0.98 &856.4 &481.2 & 3.11 \\
    \hline\hline
\end{tabular*}
\caption{Models of ALP production from Betelgeuse. The stage of stellar evolution is parametrized by the time remaining until the core collapse for Betelgeuse, $t_\mathrm{cc}$. See text for the definition of other parameters.}
\label{tab:axion_models}
\end{table*}

\noindent
\section{Introduction}
Axions and, more generally, axion-like particles (ALPs) are a prediction of several theories that attempt to complete the Standard Model of particle physics (see, e.g., Refs.~\cite{Jaeckel:2010ni,Ringwald:2014vqa,DiLuzio:2020wdo,Agrawal:2021dbo}). 
In the most generic scenarios, ALPs are very light (sub-eV) pseudoscalar particles that can be described by the effective low-energy Lagrangian
\begin{eqnarray}
    {\mathcal L}_{a}&=&\frac12 (\partial_\mu a)^2 
    -\frac12 m_a^2 a^2  \nonumber \\
   &-& \sum_{f=e,p,n} g_{af} a \bar{\psi}_f\gamma_5\psi_f
    -\frac14 g_{a\gamma} \, a\,F_{\mu\nu} \tilde{F}^{\mu\nu}\,,
    \label{Eq:L_a}
\end{eqnarray}
where $a$ is the ALP field with mass $m_a$, $\psi_f$  represent the electron and nucleon fields, and 
$F_{\mu\nu}$ and $\tilde F^{\mu\nu}$ denote the electromagnetic field strength and its dual. 
In such a description, the ALP interactions with the SM 
are parametrized by a set of  coupling constants $g_{a,i}$ $(i=f, \gamma)$. 
Particularly significant is the two photon vertex, which permits the conversion of axions into photons in an external electric or magnetic field~\cite{Raffelt:1987im}.
This phenomenon is often employed as the basis for direct ALP detection (see, e.g., Refs.~\cite{Irastorza:2018dyq,DiLuzio:2020wdo,Sikivie:2020zpn} for recent reviews).

As shown in Eq.~\eqref{Eq:L_a}, besides the axion-photon vertex, ALPs are in general expected to have non-vanishing couplings also to electrons and nucleons. 
These couplings are also currently exploited in direct detection experiments, such as CASPEr-gradient~\cite{Garcon:2017ixh,JacksonKimball:2017elr}, ARIADNE~\cite{Arvanitaki:2014dfa}, and QUAX~\cite{Barbieri:2016vwg,Crescini:2017uxs}. 
However, the benefits of the axion-fermion couplings are especially relevant in the astrophysical context, since they mediate several ALP production mechanisms in stars (see Ref.~\cite{DiLuzio:2021ysg, Giannotti:2017hny,DiVecchia:2019ejf} for review articles). 
Furthermore, the solar ALP flux includes a significant contribution from the couplings to electrons~\cite{Redondo:2013wwa,Hoof:2021mld} and nucleons~\cite{DiLuzio:2021qct}, allowing for efficient direct detection in next-generation axion helioscope searches~\cite{IAXO:2020wwp, IAXO:2019mpb}.

In this work, we turn to another astrophysical laboratory for light ALPs: Beteleguese,
a nearby ($d~{\sim}~200\mathrm{\,pc}$) red supergiant star~\cite{Harper_2008,Harper_2017}, 
with spectral type M2Iab and mass $\sim 15-24\,M_{\odot}$~\cite{luo2022stellar}. 
The Betelgeuse core is expected to be between 20 and 200 times hotter 
than the core of our Sun and, hence, provide a very promising environment to study ALPs. The most effective production mechanisms rely on the ALP coupling with photons ($g_{a\gamma}$) and electrons ($g_{ae}$), while the ALP-nucleon coupling is subdominant. For the range of couplings of interest in our discussion, once produced, such ALPs can escape the star unimpeded and convert into hard X-ray photons in the Galactic magnetic field.  

The constraints on $g_{a\gamma}$ from a 50\,ks observation of Betelgeuse using the \emph{NuSTAR} hard X-ray telescope were previously reported~\cite{Xiao:2020pra}, under the assumption that ALPs coupled only with photons.
This assumption, though valid to derive conservative constraints on the axion-photon coupling, neglected efficient ALP production channels that proceed through the coupling to electrons.
Here, we present the first complete estimates of the expected ALP flux from Betelgeuse and compare the expected
hard X-ray signal to direct observation. 

\section{ALP-photon fluxes from Betelgeuse}
\subsection{ALP Production and Betelgeuse Stellar Models}
A generic ALP, described by the Lagrangian in 
Eq.~\eqref{Eq:L_a}, can be produced in a stellar medium through processes involving photons and electrons~\footnote{
A list of processes relevant in the astrophysical contest, including useful numerical rates, can be found in the appendix of Ref.~\cite{Straniero:2019}.}. The most efficient of those involving the ALP-photon coupling is the Primakoff process, $\gamma + Ze \rightarrow a + Ze$, corresponding to the conversion of a photon into an ALP in the electrostatic field of an ion (see, e.g., Ref.~\cite{Carlson:1995wa}). The most effective ALP production rates induced by the ALP-electron coupling, in the plasma conditions typical of the Betelgeuse core, are the Compton scattering, $\gamma + e \rightarrow e + a$~\cite{Raffelt:1996wa}, and to a lesser extent, the electron-ion Bremsstrahlung, $e + Ze \rightarrow e + Ze + a$~\cite{Carenza:2021osu}. The total ALP spectrum is obtained by integrating these rates over the volume of the star. 
Alpha Orionis (Betelgeuse) is a red supergiant whose luminosity, effective temperature and metallicity are, respectively, $\log L/L_\odot=5.10\pm0.22$~\cite{Bertre:2012bh}, $T_{\rm eff}=3641\pm53\,$K~\cite{Perrin:2004ce}, and $[\mathrm{Fe/H}]=+0.1\pm0.2$~\cite{Lambert:1984}, constraining the initial mass in the range 18-22 $M_\odot$, in agreement with previous determinations~\cite{Meynet2013, Dolan2016}. This uncertainty on the mass of Betelgeuse is negligible compared to the uncertainties on the time to core collapse or the magnetic field on the line-of-sight between the detector and Betelgeuse. In the following we adopt stellar models of 20 $M_\odot$ with solar composition.

In order to model the structure of Betelgeuse, we consider stellar profiles computed by using Full Network Stellar evolution code (FuNS, see~\cite{Straniero:2019} for a detailed description of the adopted input physics and numerical algorithms of this code, where all the major uncertainties in the modeling of massive stars are extensively discussed). To estimate the ALP source spectrum, we produced 12 numerical models of Betelgeuse using the FuNS, all reproducing the observed position in the Hertzsprung–Russell diagram. The models cover a wide range of stellar evolutionary phases which reproduce the observational data, as detailed in the following. For each model we report, in Tab.~\ref{tab:axion_models}, the surface temperature and luminosity, and the time ($t_{cc}$) to core collapse. Model 0, the less evolved, represents a star in the He-burning phase, while model 11 corresponds to the Ne-burning phase. Note that, any models with $t_{cc}$ earlier than that of model 0 do not fit the observed L, $T_{\rm eff}$. In addition, comparing to Ref.~\cite{Xiao:2020pra}, we have excluded one advanced model with $t_{cc}$ less than 3.6 years since the \emph{NuSTAR} observations date back to August 2019.
\begin{figure}[t]
\centering
\includegraphics[width=\columnwidth]{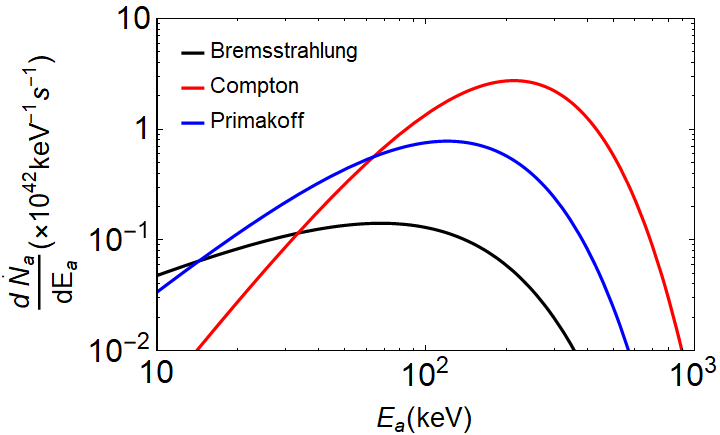}
\caption{Expected ALP fluxes from electron-ion Bremsstrahlung (black), Compton (red), and Primakoff (blue) production using $g_{ae}=10^{-13}$, $g_{a\gamma}=10^{-11}~{\rm GeV}^{-1}$, and a model of Betelgeuse with $t_{cc}$= 480 years (model 7 in Tab.~\ref{tab:axion_models}). }
\label{fig:fluxes}
\end{figure}

The specific ALP production rate from Betelgeuse, the number of emitted ALPs with energy E per unit time and volume ($\frac{\mathrm{d} \dot{n_{a}}}{\mathrm{d} E}$), is a function of the local temperature, density and chemical composition. In practice, for each of the 12 models listed in Tab.~\ref{tab:axion_models}, the ALP production rate for the Compton and Bremsstrahlung processes can be obtained by following Refs.~\cite{Raffelt:1996wa,Carenza:2021osu}, and the Primakoff production was discussed in Ref.~\cite{Xiao:2020pra}. ALPs produced through these processes have a quasi-thermal spectrum, with average energy from several 10 to several 100 keV, depending mostly on the (unknown) age of the star. These profiles have been interpolated with a cubic spline and then integrated over the whole stellar volume, then the total ALP number per time and energy can be obtained, $d\dot N_{a}/dE = \int (d\dot n_{a}/dE)dV$. Fig.~\ref{fig:fluxes} shows the expected ALP energy spectra from the electron-ion Bremsstrahlung (black line), Compton (red line) and Primakoff (blue line) processes with couplings $g_{ae}=10^{-13}$ and $g_{a\gamma}=10^{-11}~{\rm GeV}^{-1}$. The stellar model used in this example has $t_{cc}$= 480 yr, corresponding to the core C-burning stage, see Tab.~\ref{tab:axion_models}. In this specific case, we note that the dominant ALP production, through the Compton process, peaks at $\sim$O(300) keV, while the Primakoff contribution is peaked at slightly lower energies. Numerical fits for these contributions at different stages of the stellar evolution are listed in Tab.~\ref{tab:axion_models}. Despite the similarities in the surface temperature and luminosity of our stellar models, the core density and temperature grow steeply with the age of the star, increasing rapidly the ALP production rate and making the ALP spectrum harder.

With an excellent approximation, practically, the overall ALP source spectrum from Betelgeuse has the following form \cite{Andriamonje:2007ew}:
\begin{eqnarray}
    \frac{d\dot{N}_{a}}{dE} &=& \frac{10^{42}}{\textrm{keV}~\textrm{s}} \left[ C^{B}g_{13}^{2}\left(\frac{E}{E_{0}^{B}}\right)^{\beta^{B}}e^{-(\beta^{B}+1) E/E_{0}^{B}} \nonumber \right.\\
    &+&C^{C}g_{13}^{2}\left(\frac{E}{E_{0}^{C}} \right)^{\beta^{C}}e^{-(\beta^{C}+1)E/E_{0}^{C}} \nonumber \\
    &+& \left. C^{P}g_{11}^{2} \left(\frac{E}{E_{0}^{P}} \right)^{\beta^{P}}e^{-(\beta^{P}+1)E/E_{0}^{P}} \right] \, ,
    \label{eq:alp_product}
\end{eqnarray}
where $g_{11}=g_{a\gamma}/10^{-11}\,{\rm GeV}^{-1}$, $g_{13}=g_{ae}/10^{-13}$, $C^{B/C/P}$ is the normalization, $E_0^{B/C/P}$ is the average energy, and $\beta^{B/C/P}$ is the spectral index for Bremsstrahlung, Compton and Primakoff processes, respectively. The values of $C$, $E_0$ and $\beta$ for model 7 are obtained by best fitting the spectra in Fig.~\ref{fig:fluxes}, and the values for more stellar models are also reported in Tab.~\ref{tab:axion_models}.

\subsection{ALP-Photon Prediction from Betelgeuse}
\begin{figure}[!b]
    \centering
    \subfigure[~ALP-photon production from Bremsstrahlung process]{\includegraphics[width=0.495\textwidth]{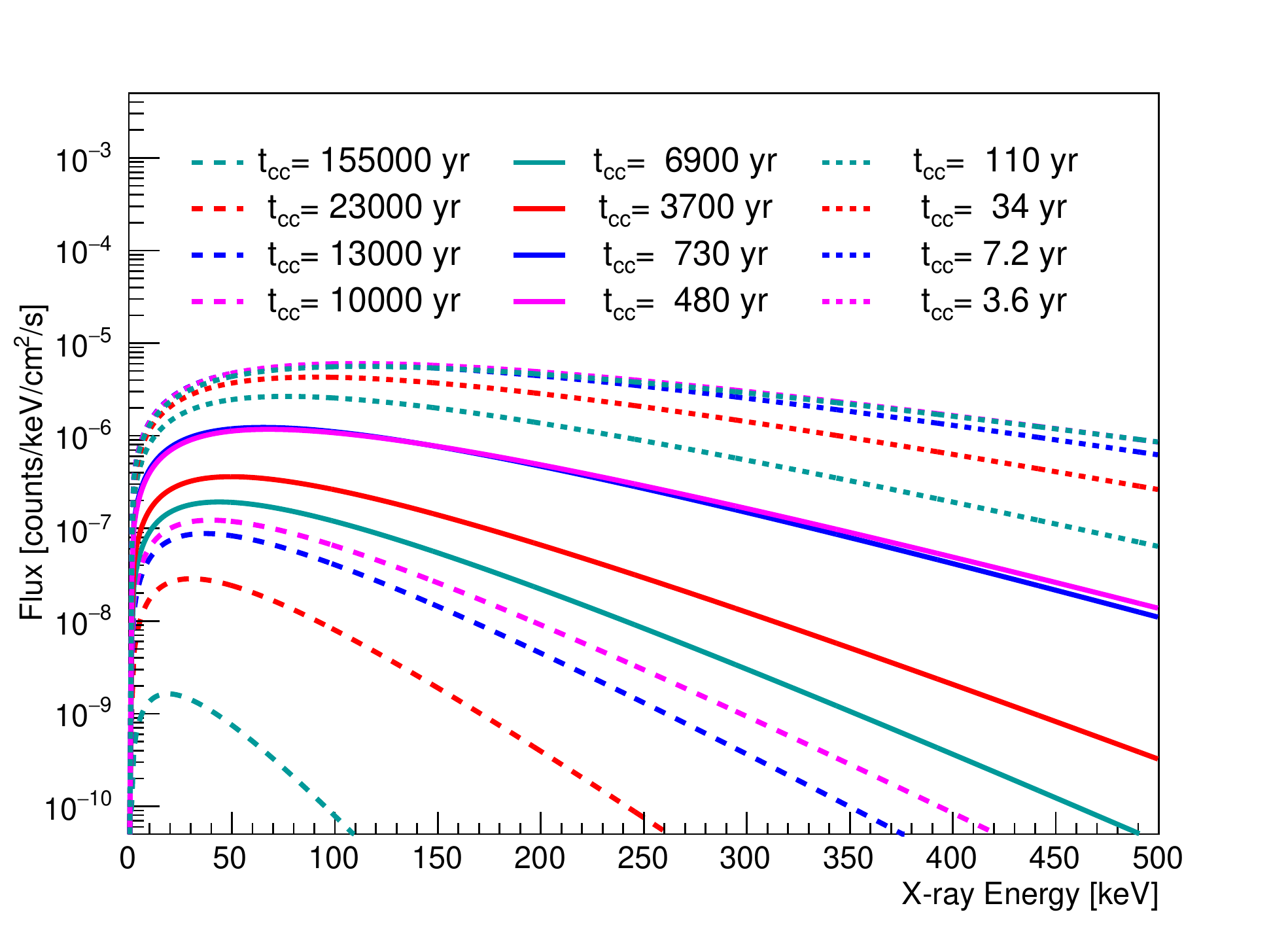}}
    \subfigure[~ALP-photon production from Compton process]{\includegraphics[width=0.495\textwidth]{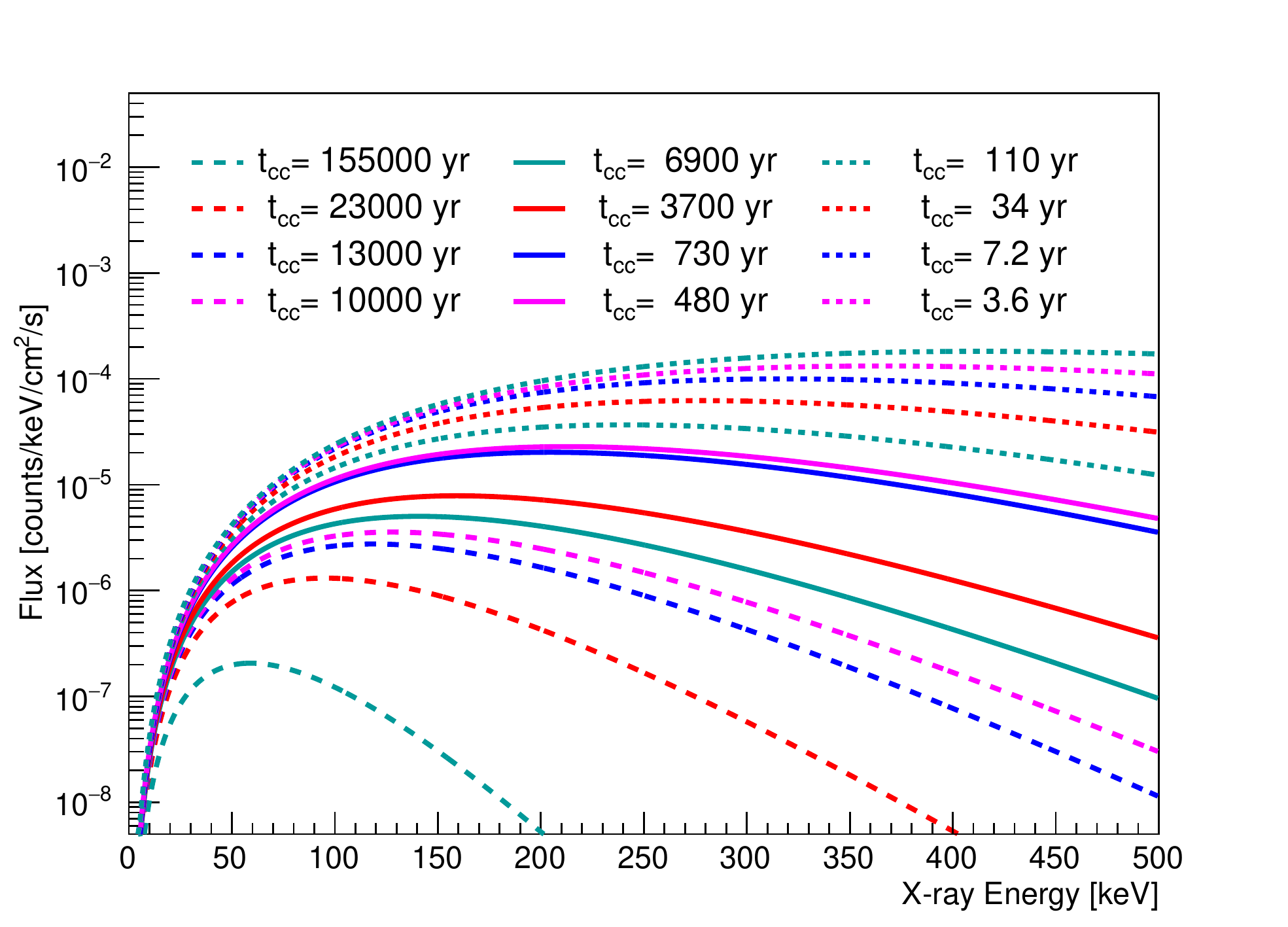}}
    \caption{Predicated $X$-ray spectra arriving at the Earth, before the instrument response, for $m_{a}=1.0\times10^{-11}$ eV, $B_{T}=1.4~\mu G$, $g_{a\gamma}=1.5\times10^{-11}~GeV^{-1}$ and $g_{ae}=1.0\times10^{-13}$.}
    \label{fig:ALP-photon production}
\end{figure}

\begin{figure*}[!ht]
\centering
\includegraphics[width=2.\columnwidth]{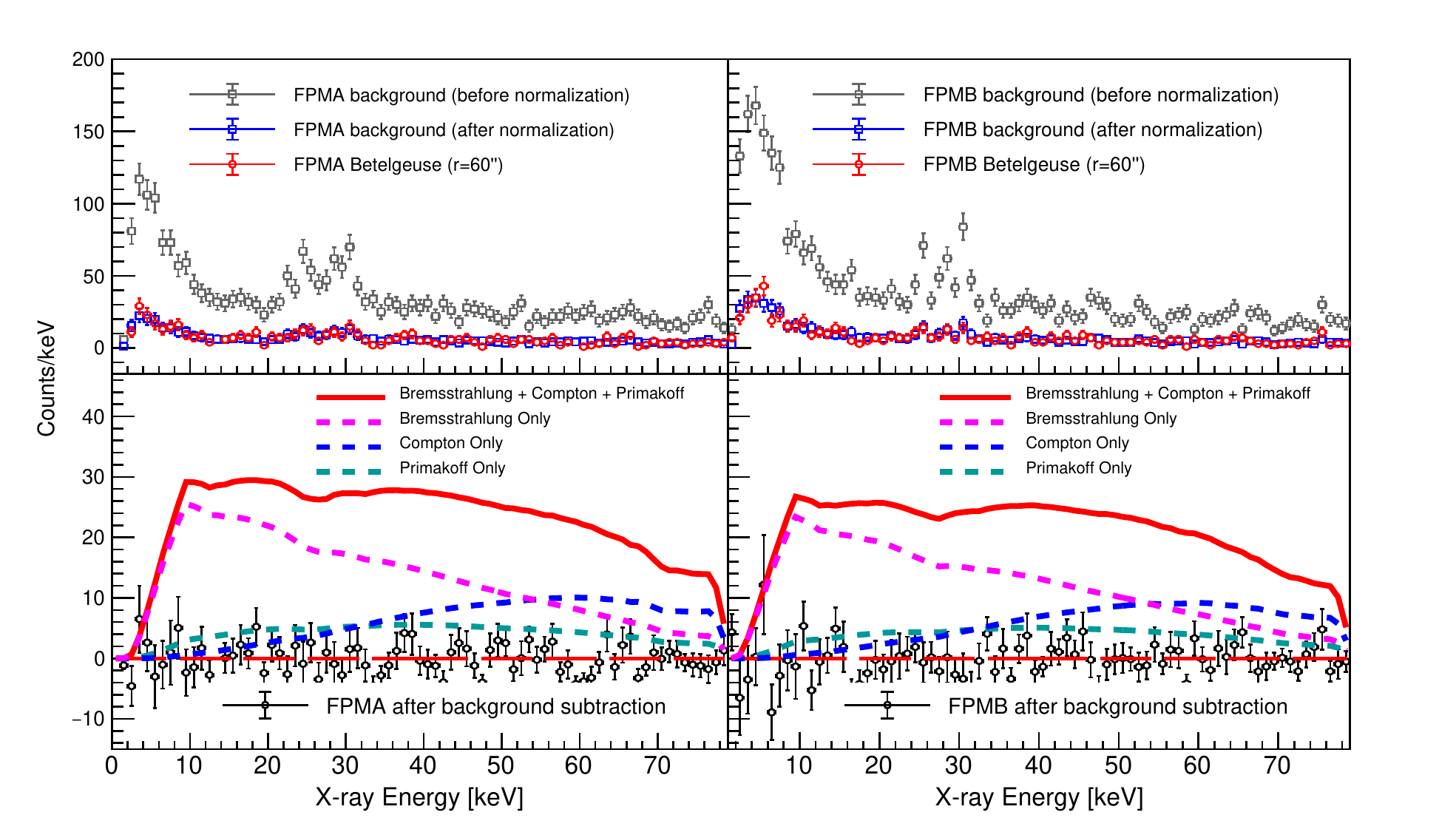}
\caption{\emph{Top:} X-ray spectra from FPMA (\emph{left}) and FPMB (\emph{right}) for the Betelgeuse source (red) and background (gray and blue for before and after normalization) regions. The error bars overlaid are the statistic uncertainties ($\sqrt{N}$). \emph{Bottom:} Source spectra after subtracting the normalized background. The errors are calculated by Sumw2 with ROOT software \cite{Brun:1997pa}. The predicted ALP-produced X-ray spectra assuming transverse magnetic field $B_{\rm T}=1.4~\,\mu\rm G$, time until core collapse $t_{cc}=$3.6~years, mass $m_a=10^{-11}\mathrm{\,eV}$, and couplings $g_{a\gamma}=1.0\times10^{-11}\mathrm{\,GeV}^{-1}$ and $g_{ae}=1.5\times10^{-13}$, that would be detected by the \emph{NuSTAR} instrument are overlaid. The stellar model parameters are described in Tab.~\ref{tab:axion_models}. The spectra are binned to a width of 1 keV, though analysis is performed on unbinned data.}
\label{fig:obs_models}
\end{figure*}

In the range of couplings we are interested in, ALPs have a negligible probability to be reabsorbed in the stellar plasma and thus leave the star unimpeded.
After leaving the star, these ALPs can convert into photons in the Galactic magnetic field~\cite{Raffelt:1987im}, causing a possibly detectable photon flux. 
Due to the relatively short distance between Betelgeuse and the Earth, $d\sim 200\,$pc, we assume that the regular component of the magnetic field ${\bf B}$ is homogeneous~\cite{Han2017,Jansson2012,XuHan2019}. 
The presence of a turbulent component on scales $\mathcal{O}$(200\,pc) or smaller~\cite{XuHan2019,Pelgrims2020,Beck2016} will not strongly affect our conclusions~\cite{Carenza:2021alz}. 

Under these assumptions, the differential photon flux per unit energy arriving at Earth is
\begin{equation}
\frac{dN_{\gamma}}{dEdSdt} = \frac{1}{4\pi d^{2}} \frac{d\dot{N_{a}}}{dE}P_{a\gamma}\,,
\label{eqa:dnr}
\end{equation}
and the ALP-photon conversion probability  is~\cite{Bassan:2010ya}
\begin{equation}
P_{a\gamma} = 8.7\times 10^{-6} g_{11}^{2}\left( \dfrac{B_\mathrm{T}}{1~\mu \mathrm{G}}\right)^2 \left( \dfrac{d}{197\,{\rm pc}}\right)^2\dfrac{\sin^2 (qd)}{(qd)^2}
\,\ ,
\label{eq:prob}
\end{equation}
where $B_\mathrm{T}$ is the transverse magnetic field, namely its component in the plane normal to the path between Earth and Betelgeuse, $d$ is the distance traveled, and $q$ is the momentum transfer, explicitly reported in Eq.~\eqref{eqa:qform}. The product of the momentum transfer $q$ and the magnetic field length $d$ can be written as
\begin{eqnarray}
    qd &\simeq& \left[77\,\left(\dfrac{m_{a}}{10^{-10}\,{\rm eV}}\right)^2-0.14\left( \dfrac{n_e}{0.013~{\rm cm}^{-3}}\right) \right] \nonumber \\ 
    &\times& \left( \dfrac{d}{197\,{\rm pc}}\right) \left(\dfrac{E}{1\mathrm{\,keV}}\right)^{-1}\,\ ,
    \label{eqa:qform}
\end{eqnarray}
Notice that for $qd\ll 1$ the conversion probability becomes energy independent and so the photon spectrum keeps the same shape of the original ALP distribution which, as discussed above,
is expected to be in the region of hard X- to soft $\gamma$-rays.
Using the parameters in Tab.~\ref{tab:axion_models}, we show the X-ray spectra arriving at Earth in Fig.~\ref{fig:ALP-photon production}.

In addition to the unknown evolutionary stage of Betelgeuse, the uncertainty in the expected photon flux at Earth is dominated by the local regular Galactic magnetic field. The reported values of the local regular magnetic field, translated to $B_T$ in the direction of Betelgeuse, vary between 0.4 $\mu$G \cite{Jansson2012} and 3.0 $\mu$G~\cite{HarveySmith:2011fe}. Here we use 1.4 $\mu$G \cite{XuHan2019} as a representative value, but note that different values will scale the expected flux -- and thus the sensitivity to couplings -- as $B_T^2$.

\section{Data Analysis}
We use a dedicated \emph{NuSTAR} observation of Betelgeuse taken on 23 August 2019 (ObsID 30501012002). Data reduction and filtering, spectral extraction, and background subtraction are identical to those in Ref.~\cite{Xiao:2020pra}. The top panel of Fig.~\ref{fig:obs_models} shows the observed X-ray spectra in the source region and background region for two independent optic and focal-plane detectors of \emph{NuSTAR}, referred to as FPMA and FPMB, respectively.

\begin{figure}[h!]
    \centering
    \subfigure[~$g_{a\gamma}=0.5\times10^{-11}~GeV^{-1}$, $g_{ae} = (0.25\sim3)\times10^{-13}$]{\includegraphics[width=0.495\textwidth]{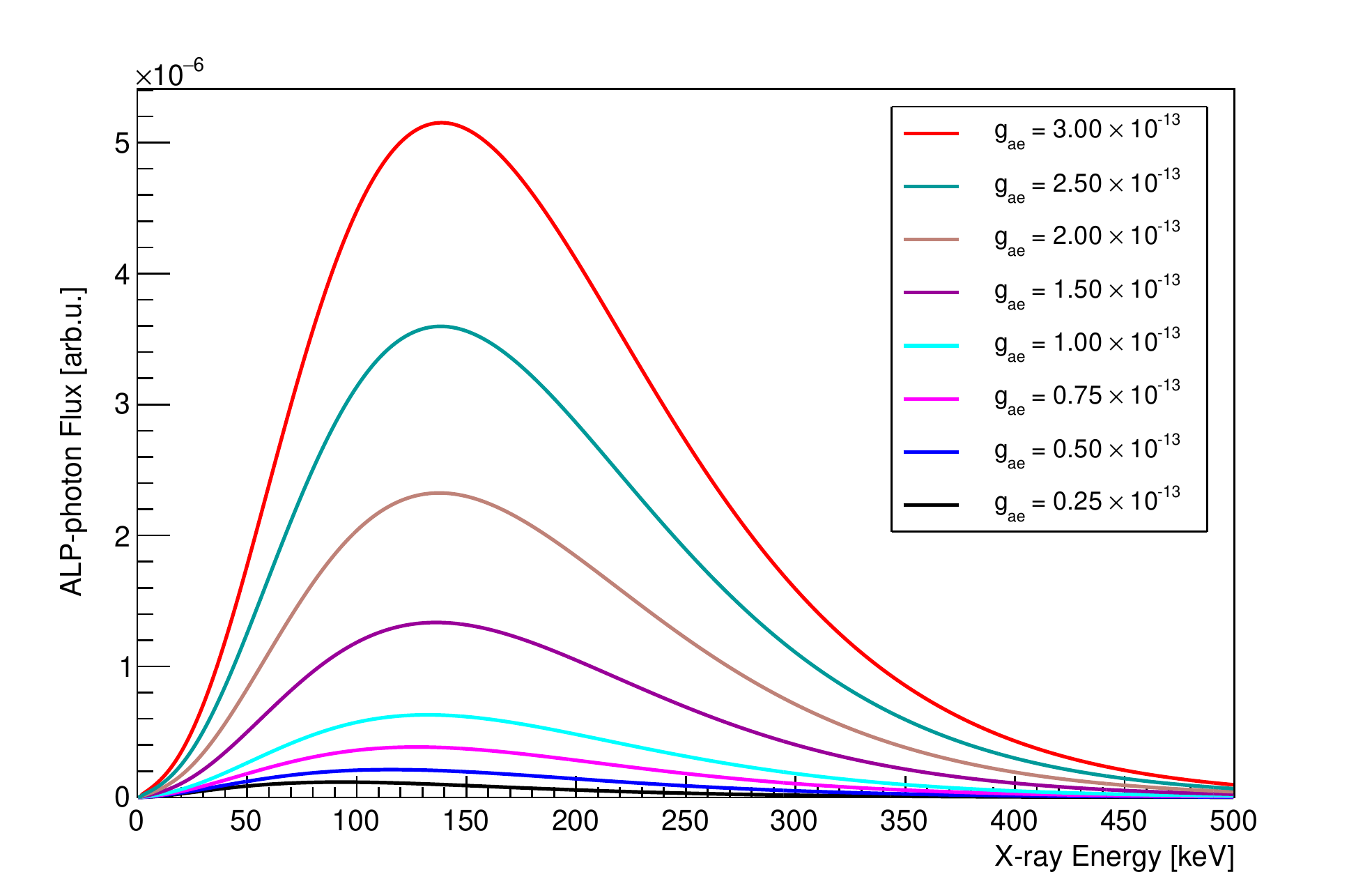}}
    \subfigure[~$g_{a\gamma}=1.0\times10^{-11}~GeV^{-1}$, $g_{ae} = (0.25\sim3)\times10^{-13}$ ]{\includegraphics[width=0.495\textwidth]{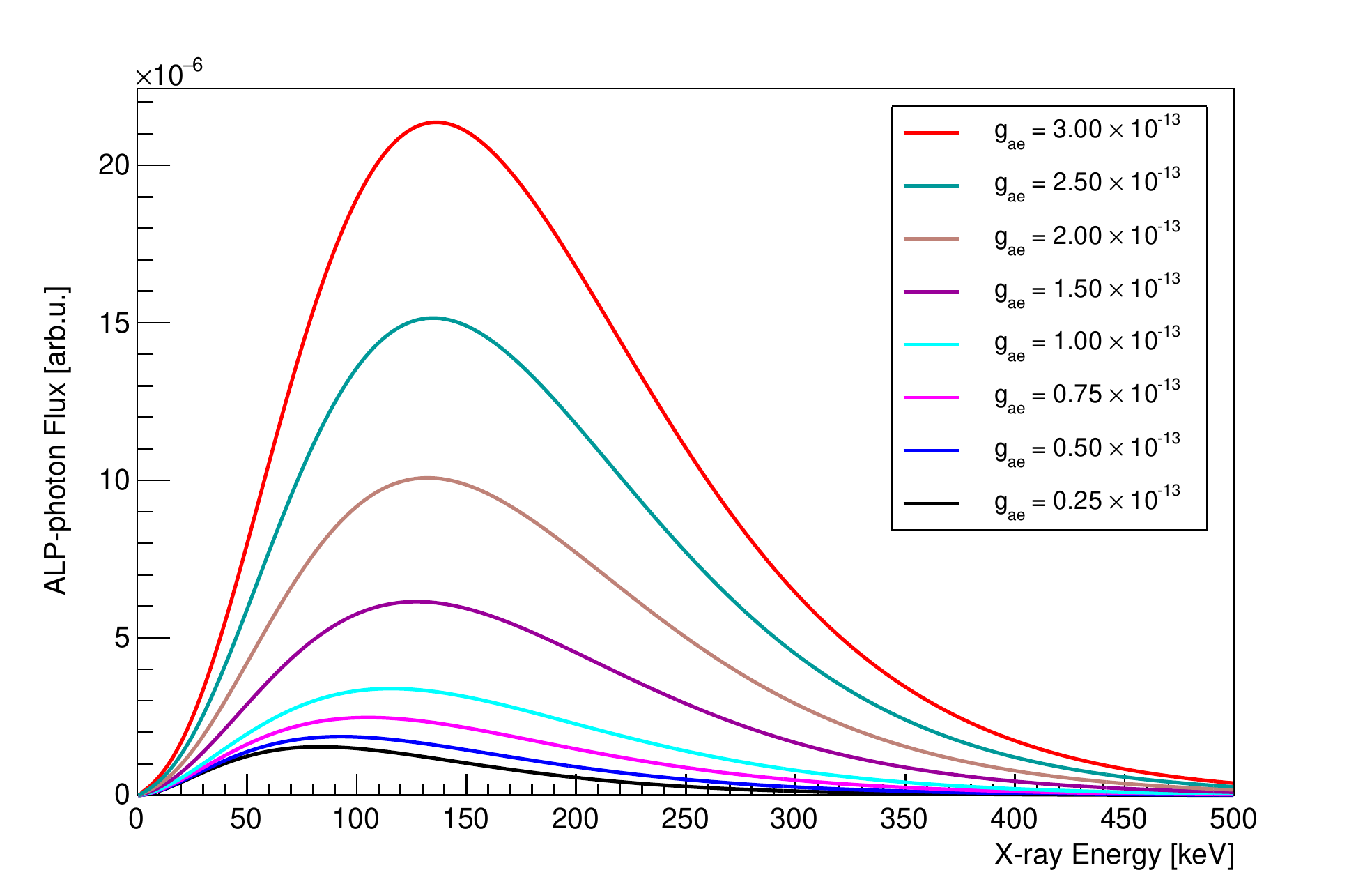}}
    \caption{Predicted X-ray spectra from the combined process of Bremsstrahlung, Compton and Primakoff after \emph{NuSTAR} instrument response for $m_{a}=1.0\times10^{-11}$ eV, $B_{T}=1.4~\mu$G, $t_{cc}$=6900 yr, and the given combinations of $g_{a\gamma}$ and $g_{ae}$.}
    \label{fig:pdf_gae_scan}
\end{figure}

An unbinned likelihood function~\cite{Junk:1999kv} is constructed as 
\begin{equation}
\mathcal{L}=\prod_{i=1}^{n}\mathcal{L}_{i} \times \prod_{i=1}^{n}\mathrm{Gauss}(\delta_\mathrm{bkg}^{i}, \sigma_\mathrm{bkg}^{i}),
\end{equation}
where n = 2 is for FPMA and FPMB with its likelihood function $\mathcal{L}_{i}$; Gauss($\delta_\mathrm{bkg}$, $\sigma_\mathrm{bkg}$) is the Gaussian penalty term with the nuisance parameter $\delta_\mathrm{bkg}$ and fractional systematic uncertainty of the background $\sigma_\mathrm{bkg}$. The likelihood function for FPMA and FPMB is constructed as
\begin{equation}
\begin{split}
\mathcal{L}_{i}=& \mathrm{Poisson}(N_\mathrm{obs}|N_\mathrm{exp}) \\
                &\times\prod_{j=1}^{N_\mathrm{obs}}\left[\frac{N_\mathrm{ax}P_\mathrm{ax}(E_\gamma^{j})}{N_\mathrm{exp}}+\frac{N_\mathrm{bkg}(1+\delta_\mathrm{bkg})P_\mathrm{bkg}(E_\gamma^{j})}{N_\mathrm{exp}}\right]
\end{split}
\end{equation}
Here, $N_\mathrm{obs}$ is the total number of events observed in our source region, and $N_\mathrm{exp} = N_\mathrm{ax}+N_\mathrm{bkg} \cdot (1+\delta_\mathrm{bkg}$) is the total number of events expected in our source region for an assumed ALP signal, where $N_\mathrm{ax}$ is the number of ALP-produced photons that would be detected by \emph{NuSTAR}; this is numerically calculated by integrating the spectrum predicated from Eq.~\eqref{eqa:dnr} in the energy range of interest after folding through the instrument response files extracted by \textsc{nuproducts}~\cite{vanLeeuwen:2007tv} for this source region. $P_\mathrm{ax}(E_\gamma)$ is the energy-dependent ALP signal probability density function (PDF), determined by $m_{a}$, $g_{ae}$, $g_{a\gamma}$, $t_\mathrm{cc}$, and $B_{T}$ (examples are shown in the lower panel of Fig.~\ref{fig:obs_models}, and more examples in a wider range are illustrated in Fig.~\ref{fig:pdf_gae_scan}). $P_\mathrm{bkg}(E_\gamma)$ is the data-driven background PDF, obtained by normalizing the background spectrum to the source region size using \textsc{nuproducts}, as described in \cite{Xiao:2020pra} and the Supplemental Material therein. $N_\mathrm{obs}$ and $N_\mathrm{bkg}$ for FPMA and FPMB are listed in Tab.1 of Ref.~\cite{Xiao:2020pra}. In the energy range of 10--79 keV this analysis is using, we observed 384 events from FPMA while expecting 393 background events, and 433 events from FPMB while expecting 441 background events. Given the statistics of expected background events in the observation region, $\sigma_\mathrm{bkg}$ is conservatively set at 10\% for both FPMA and FPMB, but allowed with independent Gaussian fluctuation in this analysis.

The standard profile likelihood test statistic \cite{Cowan:2010js, Feldman:1997qc} is used to derive constraints on the ALP coupling to electrons $g_{ae}$ and photons $g_{a\gamma}$:
\begin{equation}
q(g_\mathrm{test})=\left\{
\begin{aligned}
&-2\ln\frac{\mathcal{L}_\mathrm{max}(g_\mathrm{test}, \dot{\theta})} {\mathcal{L}_\mathrm{max}(g_\mathrm{best}, \hat{\theta})}, &g_\mathrm{test}\geq g_\mathrm{best} \\
&~0, &g_\mathrm{test}< g_\mathrm{best}
\end{aligned}
\right.
\end{equation}
where $g_{test}$ is the tested $g_{ae}$ for a given $g_{a\gamma}$ or the tested $g_{ae}\times g_{a\gamma}$ in the later analysis scenario, respectively; and $\theta$ represents the nuisance parameters which are
all allowed to vary in the fitting.

\section{Results}
\par \noindent \textit{Constraints on $g_{ae}$ v.s. $g_{a\gamma}$}---
We consider the ALP production from Betelgeuse due to the ALP coupling with photons and electrons, meaning via the combined processes of Bremsstrahlung, Compton, and Primakoff (i.e. the BCP effect). 
Combining Eqs.~\eqref{eq:alp_product} and \eqref{eq:prob}, the ALP-photon production from Betelgeuse can be formally written as:
\begin{equation}
F_{BCP} = \left[(\textrm{B} +\textrm{C}) \cdot g_{ae}^{2} + \textrm{P} \cdot g_{a\gamma}^{2}\right] \times g_{a\gamma}^{2} \cdot S_{m_{a}} \cdot B_{T}^{2},
\label{eq:alp_photon_bcp}
\end{equation}
where B, C and P are t$_{cc}$-dependent coefficients. For the light ALPs ($m_{a}\leq3.5\times10^{-11}$ eV), $S_{m_{a}}$ is no longer sensitive to the ALP masses. In this analysis, we present results for low-mass ALP, $m_{a}\leq3.5\times10^{-11} eV$. For a given $t_\mathrm{cc}$, the shape of ALP signal PDF, $P_\mathrm{ax}(E_\gamma)$, is determined by the combination of $g_{ae}$ and $g_{a\gamma}$, as illustrated in Fig.~\ref{fig:pdf_gae_scan}.

\begin{figure}[h]
\centering
\includegraphics[width=\columnwidth]{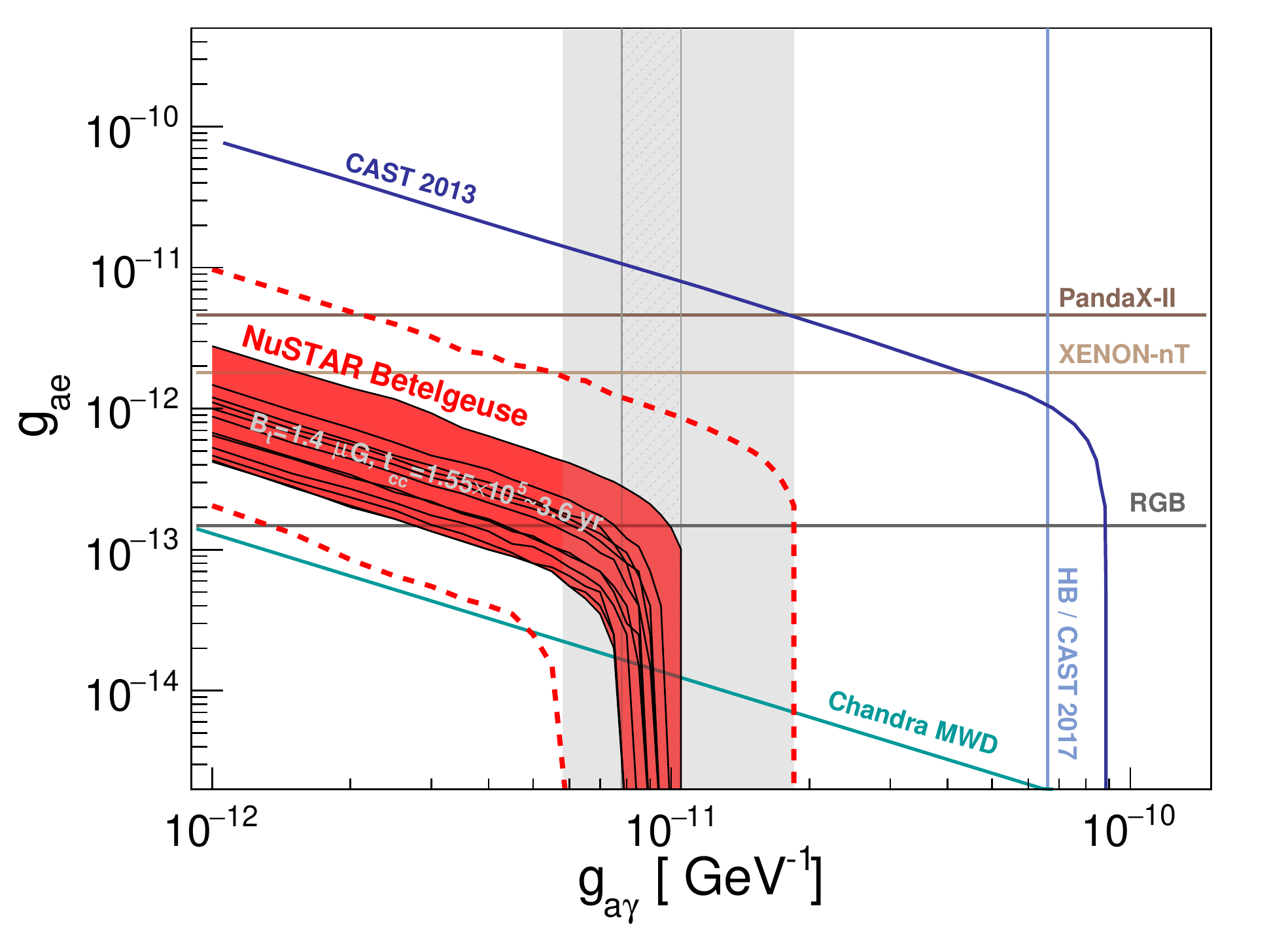}
\caption{The 95\% C.L. upper limit on $g_{ae}$ as a function of $g_{a\gamma}$ for $m_{a}\leq3.5\times10^{-11}~\textrm{eV}$. The solid black lines show the upper limit for each stellar model assuming a representative value of $B_T$=1.4~\,$\mu$G~\cite{XuHan2019}, with the red band indicating the uncertainty due to this unknown evolutionary state; the dashed red lines show the upper limit for the most conservative ($B_{T}$=0.4~\,$\mu$G and $t_{cc}$=1.55$\times10^{5}$ yrs) and most optimistic case ($B_{T}$=3.0~\,$\mu$G and $t_{cc}$=3.6 yrs). Overlaid are the bounds from the CAST experiment~\cite{Barth:2013sma} (95\% C.L.), PandaX-II complete data~\cite{PandaX-II:2020udv} (90\% C.L.), XENONnT~\cite{XENON:2022mpc}, red-giant branch (RGB) observations~\cite{Straniero:2020iyi,Capozzi:2020cbu} (95\% C.L.), and Chandra observations of magnetic white dwarf~\cite{Dessert:2021bkv} (95\% C.L.). 
The region of $g_{a\gamma}$ excluded by our previous analysis~\cite{Xiao:2020pra} is labeled, with the hatched band indicating the uncertainty due to stellar modeling. The constraints on $g_{a\gamma}$ from CAST latest results~\cite{cast:2017ftl} and horizontal branch (HB) stars in globular clusters ~\cite{Ayala:2014pea, Straniero:2015nvc} are also indicated.
}
\label{fig:results_BCandP}
\end{figure}

For $g_{a\gamma}$ in the range from $10^{-12}$ $\textrm{GeV}^{-1}$ to the upper band that has been set in Ref.~\cite{Xiao:2020pra}, where the same data set was analyzed by assuming only Primakoff ALP production from Betelgeuse, we scan through the ALP-electron coupling, $g_\mathrm{test}$, and perform two maximum likelihood fits: one with the $g_{ae}$ as its best fit value, $g_\mathrm{best}$, and the other with $g_{ae}$ fixed at $g_\mathrm{test}$. 
The nuisance parameters are all allowed to vary in both to achieve the best fit. Assuming $q(g_\mathrm{test}$) follows a half-$\chi^{2}$ distribution with a single degree of freedom \cite{Cowan:2010js}, we derive the 95\% C.L. upper limit on $g_{ae}$ for a given $g_\mathrm{test}$. 

Constraints on $g_{ae}$ as a function of $g_{a\gamma}$ are shown in Fig.~\ref{fig:results_BCandP} for $m_{a} \leq 3.5\times 10^{-11}$ eV and for the 12 modeled stellar stages, $t_{cc}$ from 3.6 yrs to 1.55$\times10^{5}$ yrs, with the representative Galactic magnetic field, $B_{T}$=1.4 $\mu G$. We also show the results for the most conservative case ($B_{T}$=0.4~\,$\mu$G and $t_{cc}$=1.55$\times10^{5}$ yrs) and most optimistic case ($B_{T}$=3.0~\,$\mu$G and $t_{cc}$=3.6 yrs).
When $g_{a\gamma}$ reaches the region around $1\times10^{-11}$ GeV$^{-1}$, the Primakoff process dominates the ALP production, and a tight limit is set on $g_{ae}$. 
This is consistent with the analysis in the previous work~\cite{Xiao:2020pra}, shown as the hatched gray band in Fig.~\ref{fig:results_BCandP}, where $g_{ae}$ was assumed to be zero and the band width corresponds to the uncertainties due to stellar modeling.

\begin{figure}[t]
\centering
\includegraphics[width=\columnwidth]{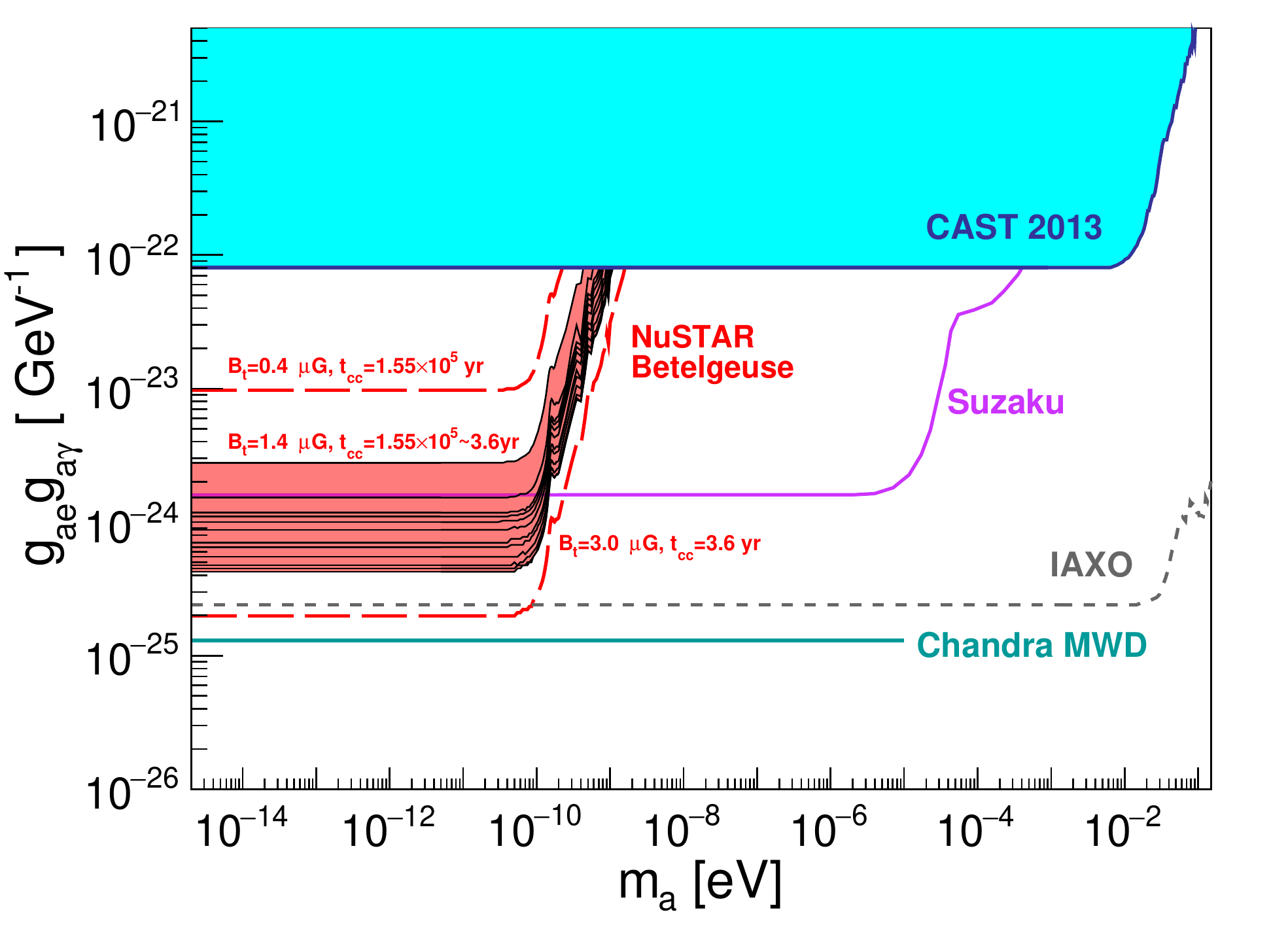}
\caption{
The 95\% C.L. upper limits of $g_{ae} \times g_{a\gamma}$ as a function of ALP mass. The solid black lines show the upper limit for each stellar model, assuming a representative value of $B_T$=1.4\,$\mu$G~\cite{XuHan2019}, with the red band indicating the uncertainty due to this unknown evolutionary state. The constraints will scale with different $B_T$ as in Eq.~\eqref{eq:prob}, the two dashed red lines show the upper limit for the most conservative ($B_{T}$=0.4~\,$\mu$G and $t_{cc}$=1.55$\times10^{5}$ yrs) and most optimistic case ($B_{T}$=3.0~\,$\mu$G and $t_{cc}$=3.6 yrs). Overlaid are the limit from CAST~\cite{Barth:2013sma} and the projected sensitivity of IAXO~\cite{IAXO:2020wwp}, as well as the limits from Suzaku~\cite{Dessert:2019sgw} and Chandra observations of magnetic white dwarfs~\cite{Dessert:2021bkv}.}
\label{fig:results_BandC}
\end{figure}

This limit on $g_{ae}$ is almost two orders of magnitude stronger than that placed by the CAST experiment using solar axions~\cite{Barth:2013sma}, for 
$g_{a\gamma} \gtrsim 1\times10^{-12}$ GeV$^{-1}$.
Although our derived limit scales with the assumed $B_T^2$, even the most conservative value of $B_T$=0.4~\,$\mu$G~\cite{Jansson2012} gives a limit that is a factor of (8$\sim$50) (depending on the stellar model) stronger than CAST. 
For the representative $B_T$=1.4~\,$\mu$G, our limit also supersedes the one placed by the Particle and Astrophysical Xenon experiment (PandaX)~\cite{PandaX-II:2020udv} (90\% C.L.) and XENONnT's latest results~\cite{XENON:2022mpc} (90\% C.L.). On the other hand, the recent Chandra observation of magnetic white dwarfs~\cite{Dessert:2021bkv} supersedes our bound for $B_T$=1.4~\,$\mu$G by a factor of (3$\sim$20) (depending on the Betelgeuse stellar model), but is very close to the results from our most optimistic stellar model and $B_T$. The astrophysical bound from energy loss in red-giant branch stars also gives a stronger constraint, $g_{ae} < 1.48 \times 10^{-13}$~\cite{Straniero:2020iyi, Capozzi:2020cbu}. 
However, for $g_{a\gamma} \sim 10^{-11}~\textrm{GeV}^{-1}$, the Betelgeuse bound on $g_{ae}$ supersedes this red-giant bound. 

\par \noindent \textit{Constraints on $g_{ae}$$\times$$g_{a\gamma}$}---
For small values of $g_{a\gamma}$, the Primakoff emission from Betelgeuse is subdominant, and therefore the Bremsstrahlung and Compton process discussed in this work dominate the ALP production. 
By ignoring the Primakoff process, the ALP production rate is scaled with $g_{ae} \times g_{a\gamma}$. In this scenario, we are able to profile out $g_{ae} \times g_{a\gamma}$ in the likelihood function and perform the fitting on the product of $g_{ae} \times g_{a\gamma}$ as a function of $m_a$ for given $t_\mathrm{cc}$ and $B_{T}$.

The resulting 95\% C.L. bound is shown in Fig.~\ref{fig:results_BandC}, with the red band for $B_{T}$=1.4~\,$\mu$G, indicating the uncertainty due to stellar model, and two dashed red lines showing the most conservative case ($B_{T}$=0.4~\,$\mu$G and $t_{cc}$=1.55$\times10^{5}$ yrs) and most optimistic case ($B_{T}$=3.0~\,$\mu$G and $t_{cc}$=3.6 yrs). For very light ALPs, the upper limits of $g_{ae} \times g_{a\gamma}$ are constant, which are responsible for the region where there is an anti-correlation between $g_{ae}$ and $g_{a\gamma}$ in Fig.~\ref{fig:results_BCandP}. 
We note that this analysis provides conservative constraints on $g_{ae} \times g_{a\gamma}$ when $g_{a\gamma}$ is small enough that the Primakoff contribution to ALP production can be ignored. However, as already discussed previously, once the Primakoff process is more pronounced, much more stringent constraints on $g_{ae}$ can be set, and then the band of $g_{ae} \times g_{a\gamma}$ would move lower accordingly. 
This is reflected by the slopes of the lines in Fig.~\ref{fig:results_BCandP}.

For $m_a \leq (3.5-5.5) \times 10^{-11}$eV (depending on the stellar model), this Betelgeuse bound for
the representative Galactic magnetic field, $B_T$=1.4~\,$\mu$G, is $\sim$1.5 orders of magnitude stronger than the solar ALP bound from CAST~\cite{Barth:2013sma}. In the same mass range, our bound is comparable or even stronger (depending on the stage of the stellar evolution and assumed $B_T$) than that derived by the non-observation by Suzaku of X-rays from ALP conversions in magnetic white dwarf stars~\cite{Dessert:2019sgw}.
Only the recent  Chandra observation of magnetic white dwarfs~\cite{Dessert:2021bkv} supersedes our bound by a factor $\gtrsim 5$ for
the representative Galactic magnetic field, $B_T$=1.4~\,$\mu$G, but is very close to our most optimistic case.

\section{Conclusions and Discussion}
In this study we have presented, for the first time, the constraints on ALPs coupled to both electrons and photons from a dedicated \emph{NuSTAR} observation of Betelgeuse. 
Light ALPs are efficiently produced in the hot core of this supergiant star, mostly through Primakoff, Bremsstrahlung, and Compton processes, and transformed into a hard X-ray flux in the Galactic magnetic field.
Previous work reported the derived limits on the ALP-photon coupling only, conservatively assuming only Primakoff process production in Betelgeuse~\cite{Xiao:2020pra}.
This new work allows stringent bounds on the combined ALP-photon and ALP-electron couplings.

Our limit on $g_{ae}$ as a function of $g_{a\gamma}$ for the case of $m_a \leq 3.5\times10^{-11}$eV supersedes the limits from CAST, PandaX-II, and XENONnT in some regions of the parameter space. For the assumed value of $B_T$, the parameter region probed extends below the strong constraints from red-giant branch and horizontal branch stars, and partially covers the area hinted by the cooling of horizontal branch stars~\cite{Ayala:2014pea,Straniero:2015nvc}. 

In the case of very light ALPs, $m_a \leq (3.5-5.5) \times 10^{-11}$eV (depending on the stellar model), our constraint on $g_{ae} \times g_{a\gamma}$ is among the most stringent, improving by $\sim$1.5 orders of magnitude the bound on solar ALPs from CAST. Presently, only the study of conversions in magnetic white dwarfs provide a more stringent bounds. Our analysis offers an independent support to the exclusion of this region of the ALP parameter space. 
Finally, our analysis can be extended essentially unchanged to other close-by supergiant stars~\cite{Mukhopadhyay:2020ubs}, providing new possibilities to study this region of the ALP parameter space.



\section{Acknowledgements}
We thank D. R. Wik and S. Rossland for helpful discussions on the \emph{NuSTAR} instrument background;  J.L. Han and S. Zhang for discussions concerning the Galactic magnetic field in the direction of Betelgeuse. The \emph{NuSTAR} observations described in this work were awarded under NASA Grant No. 80NSSC20K0031. The computational aspects of this work made extensive use of the following packages: \textsc{saoimage ds9} distributed by the Smithsonian Astrophysical Observatory; the \textsc{scipy} ecosystem, particularly \textsc{matplotlib} and \textsc{numpy}; and \textsc{astropy}, a community-developed core \textsc{python} package for Astronomy. This research has made use of data and software provided by the High Energy Astrophysics Science Archive Research Center (HEASARC), which is a service of the Astrophysics Science Division at NASA/GSFC and the High Energy Astrophysics Division of the Smithsonian Astrophysical Observatory. We thank the \emph{NuSTAR} Operations, Software and Calibration teams for support with the execution and analysis of these observations. This research made use of the \emph{NuSTAR} Data Analysis Software (NuSTARDAS), jointly developed by the ASI Science Data Center (ASDC, Italy) and the California Institute of Technology (USA). M.X. and M.G. were supported by NASA Grant No. 80NSSC20K0031. O.S. has been supported by the Agenzia Spaziale Italiana (ASI) and the Instituto Nazionale di Astrofsica (INAF) under the agreement  n.   2017-14-H.0  - attivit\`a  di  studio  per  la  comunit\`a  scientifica  di  Astrofisica  delle  Alte Energie  e  Fisica Astroparticellare. A.M. is partially supported by the Italian Istituto Nazionale di Fisica Nucleare (INFN) through the ``Theoretical Astroparticle Physics'' project and by the research grant number 2017W4HA7S ``NAT-NET: Neutrino and Astroparticle Theory Network'' under the program PRIN 2017 funded by the Italian Ministero dell'Universit\`a e della Ricerca (MUR). B.G. was supported under NASA contract NNG08FD60C.
The work of P.C. is supported by the European Research Council under Grant No. 742104 and by the Swedish Research Council (VR) under grants  2018-03641 and 2019-02337.

\bibliography{Betelgeuse}

\end{document}